\documentclass[a4paper]{llncs}
\usepackage{amsmath,amssymb,amsfonts}
\usepackage{amssymb}
\usepackage{graphicx}
\usepackage{amsfonts}
\usepackage{subfigure}
\usepackage{graphicx}
\usepackage{algorithm}
\usepackage{algorithmic}
\usepackage{amsmath}

\usepackage{url}
\urldef{\mailsa}\path|{alfred.hofmann, ursula.barth, ingrid.haas, frank.holzwarth,|
\urldef{\mailsb}\path|anna.kramer, leonie.kunz, christine.reiss, nicole.sator,|
\urldef{\mailsc}\path|erika.siebert-cole, peter.strasser, lncs}@springer.com|    
\newcommand{\keywords}[1]{\par\addvspace\baselineskip
\noindent\keywordname\enspace\ignorespaces#1}

\makeatletter 
\renewcommand{\thefigure}{%
\thesection.\arabic{figure}} 
\@addtoreset{figure}{section} 
\makeatother 

\newcommand{\bi}[1]{\ensuremath{\boldsymbol{#1}}}

\newcommand{\wt}[1]{\ensuremath{\widetilde{#1}}}

\def\rnum#1{\expandafter{\romannumeral #1}}

\newcommand{\num}{\ensuremath{\gamma}}

\begin{document}

\mainmatter  

\title{Recommendation with $k$-anonymized Ratings}

\titlerunning{Recommendation with $k$-anonymized Ratings}

%
%
%


\author{Jun Sakuma\inst{1} \and Tatsuya Osame\inst{1}}
\institute{University of Tsukuba, 1-1-1 Tennodai, Tsukuba, Ibaraki, 305-8577 Japan\\
\email{jun@mcs.tsukuba.ac.jp} and \email{tatsuya@cs.tsukuba.ac.jp}
}

%
%

\toctitle{Recommendation with $k$-anonymized Ratings}
\tocauthor{Jun Sakuma and Tatsuya Osame}
\maketitle

\begin{abstract}
Recommender systems are widely used to predict personalized preferences of goods or services using users' past activities, such as item ratings or purchase histories. 
If collections of such personal activities were made publicly available, they could be used to personalize a diverse range of services, including targeted advertisement or recommendations. However, there would be an accompanying risk of privacy violations. 
The pioneering work of Narayanan et al.\ demonstrated that even if the identifiers are eliminated, the public release of user ratings can allow for the identification of users by those who have only a small amount of data on the users' past ratings. 

In this paper, we assume the following setting. A collector collects user ratings, then anonymizes and distributes them. A recommender constructs a recommender system based on the anonymized ratings provided by the collector.
Based on this setting, we exhaustively list the models of recommender systems that use anonymized ratings. 
For each model, we then present an item-based collaborative filtering algorithm for making recommendations based on anonymized ratings. 
Our experimental results show that an item-based collaborative filtering based on anonymized ratings can perform better than collaborative filterings based on 5--10 non-anonymized ratings. 
This surprising result indicates that, in some settings, privacy protection does not necessarily reduce the usefulness of recommendations.
From experimental analysis of this counterintuitive result, we observed that the sparsity of the ratings can be reduced by anonymization and the variance of the prediction can be reduced if $k$, the anonymization parameter, is appropriately tuned. In this way, the predictive performance of recommendations based on anonymized ratings can be improved in some settings.
\end{abstract}
\section{Introduction}

With rapid advances in online services, a huge amount of information describing users' personal activities is being collected and stored. Recommender systems that use records of user activities are widely used in diverse services, including e-commerce and information dissemination. On the other hand, records describing detailed user activities can sometime compromise privacy.
For example, Narayanan et al.\ have reported an anonymity assessment of a dataset containing user ratings of movies, which was published as part of a competition for recommendation algorithms~\cite{Narayanan2008}. 
They demonstrated that even if the identifiers (i.e., the names of the users) are eliminated from the dataset, the public release of the dataset can allow for the reidentification of users.
For example, 84\% of users can be uniquely reidentified using only their preferences for eight movies that are not among the top 500 most popular movies. 
Due to the sparsity of user ratings, such reidentification can happen even with only a small amount of background knowledge about a target.

Considering that user ratings for recommendation are always sparse, some privacy protection is needed when user ratings are released. 
In this paper, we consider practical models for making recommendations based on anonymized ratings.
We then introduce recommendation algorithms that fit these models and achieve a good balance of usefulness and preservation of privacy. 

\subsubsection{Related Works}

Recently, there have been two lines of active research concerning the preservation of privacy in recommendation systems.
The objective of one of these lines of study is to preserve the privacy of the user ratings from the entity who provides the recommendation.
Canny has presented a privacy-preserving recommender system that uses cryptographic tools and assumes that users communicate with each other over a peer-to-peer network\cite{canny2002collaborative}. 
In this system, the ratings are homomorphically encrypted, and then the matrix is completed without decryption by means of singular value decomposition. Polat et al.\ presented a recommendation scheme in which user ratings are randomized by the users themselves, and the entity that collects the user ratings performs the recommendations based on the randomized user ratings\cite{polat2003privacy}. Through experiments, they examined the trade-offs between privacy and prediction accuracy in their recommendation scheme. Parra-Arnau et al.\ introduced a recommendation system in which the users can suppress sensitive ratings \cite{parra2012privacy}. They also discussed the optimal trade-off between the suppression of ratings and privacy, but they did not explicitly discuss the effect of the privacy protection mechanism on the prediction accuracy.

In the other main line of research, the objective is to preserve privacy when releasing user ratings to be used for recommendations. 
Parameswaran et al.\ presented a scheme that first obfuscates the user ratings and then distributes them. Their obfuscation scheme does not necessarily give a mathematical guarantee of anonymity, but their experimental results revealed that it does not degrade the prediction accuracy of the recommendations\cite{parameswaran2007privacy}.
Chang et al.\ proposed an algorithm that $k$-anonymizes user ratings and then constructs a recommender system based on the anonymized ratings. Recognizing that the rating matrices are sparse, Chang et al.'s idea is to first make the rating matrix dense by complementing it using singular value decomposition, and then to $k$-anonymize the resulting matrix\cite{chang2010towards}.
Experimental results with Netflix data show that the loss of accuracy in the predictions is less with the predictive anonymization scheme than it is with the regular $k$-anonymization scheme.

\subsubsection{Our Contribution}

Assume a collector of user ratings distributes $k$-anonymized ratings to the public, such as considered in \cite{parameswaran2007privacy} and \cite{chang2010towards}. The objective of our study is make a survey of all recommender systems that use anonymized ratings and then to present item-based collaborative filtering recommendation algorithms for each of these models. 
In general, a system that provides recommendations needs to have some information associated with each of the users in order to personalize the recommendations. 
Such information is referred to as the {\em prediction input}. 
Existing privacy-preserving recommendation algorithms implicitly assume that the system is able to relate the user who obtains a recommendation to the user who provided the anonymized ratings; however, this is not always the case in practice, because there is no guarantee that the ratings of any individual user are contained in the anonymized rating matrix that the system possesses. 
Even if they are contained, the recommender cannot know this. 

Considering the above, we will introduce models for providing recommendations that are based on anonymized rating matrices, and which are determined by several input parameters that relate the individual who provided the ratings with the one who obtains the recommendation. 
We then present a recommendation algorithm for each of these models.
In some models, our experimental results show that the prediction accuracy decreases when there is a strong guarantee of anonymity. 
This indicates that there is a trade-off between anonymity and utility; this does not contradict the conclusions of \cite{chang2010towards}.
On the other hand, for some of our models, we found through experiment that the prediction accuracy of the recommendations that were based on anonymized ratings were better than those based on non-anonymized ratings, provided that the non-anonymized ratings were used as the prediction inputs.
This surprising result suggests that the guarantee of anonymity does not necessarily degrade the utility if the recommendation model is appropriately chosen. Below, we analyze why we obtained this counterintuitive result. 

\section{Preliminaries}\label{sec:definition}

\subsubsection{Recommendation System}

Let $(r_{ui}) = \bi{R} \in \mathbb{R}^{n\times m}$ be a sparse rating matrix of $n$ users and $m$ items, where $r_{ui}$ is the rating of user $u$ for item $i$. We denote the rating matrix as $(\bi{r}_1 \bi{r}_2 \hdots \bi{r}_n)^T=\bi{R}$, where $\bi{r}_u \in \mathbb{R}^m$ contains the ratings of user $u$. $\bi{R}$ contains missing values. Where $r_{ui}$ has a value in $\bi R$ as $\Omega$, we denote a set of index pairs $(u, i)$. 

Let $f: U \times I \mapsto \bi R$ be a prediction function of ratings where $U = \{1, 2, \cdots, n \}$ and $I = \{1, 2, \cdots, m \}$ denote the sets of user and item identities, respectively.
For $(u,i) \notin \Omega$, the prediction of the rating is given by $\hat{r}_{ui} =f(u,i)$.
Similarity-based collaborative filtering~\cite{herlocker1999algorithmic,Sarwar2001} and matrix completion via low-rank matrix approximation~\cite{paterek2007improving} are well known as methods for recommender systems. 
Note that although our proposal is demonstrated with an item-similarity based collaborative filtering system \cite{Sarwar2001}, the discussion in this paper is not necessarily dependent on a specific algorithm. The construction of our method, based on a low-rank matrix approximation, is presented in a longer version of this paper.

To evaluate the prediction accuracy of the recommender system, we divided $\Omega$ into two disjoint sets: a training set $\Omega_{\mbox{train}}$ and a test set $\Omega_{\mbox{test}}$. 
The set $\Omega_{\mbox{train}}$ is used to train the prediction function, and the set $\Omega_{\mbox{test}}$ is used to evaluate its accuracy at making predictions.
In our experiments, the root-mean-square error (RMSE) evaluated with $k$-fold cross-validation was used to measure the prediction accuracy of $f$.

\subsubsection{$k$-anonymity and $k$-anonymization of Rating Matrices}

Let $T$ be a database with $n$ records.
We assume each record contains information associated with a single individual and consists of the {\em identifier}, {\em quasi-identifiers} (QIs), and {\em sensitive attribute values}.
The identifier is a unique key that identifies the individual.
the QIs are user attributes, such as age or gender; the particular combination of QIs can identify a single individual with high probability.
The sensitive attribute values are the other private information.
After the identifiers have been eliminated, $T$ is said to be {\em $k$-anonymous} if, for any combination of the QIs that appear in $T$, there exist at least $k$ individuals who have that particular combination of QIs\cite{sweeney2002k}.
A set of records that share the same combination of QIs is referred to as an {\em equivalence class}.
Thus, if the size of all the equivalence classes in $T$ is greater than $k$, $T$ is $k$-anonymous.

In a recommendation system, each record contains the rating values associated with a particular individual.
The rating values are usually interpreted as sensitive attribute values because they are not user attributes. However, as already mentioned in the introductory section, in some cases, there is a high probability that individuals can be uniquely identified by the ratings, particularly when $T$ is sparse. Thus, we regard the rating values as QIs.

Let $t_u$ be a record associated with user $u$.
In our setting, a record represents a rating vector,  $t_u = \bi r_u \in{ \bf R}^m$, and the database table is given as $T=\{ {\bi r}_u \}_{u=1}^n$.
Following the definition of $k$-anonymity\cite{sweeney2002k}, for any rating vector in $T$, if there exist at least $k-1$ other rating vectors having the same rating values, then the database is $k$-anonymous.
Let $ \wt{T}=\{ ( \wt{\bi r}_u, k_u)\}_{u=1}^{n'} $ represent the $k$-anonymized database of $T$. 
Here, $\wt{\bi r}_u$ and $k_u$ mean that the number of rating vectors $\wt{\bi r}_u$ contained in $T$ is $k_u$. Since $\wt{T}$ is $k$-anonymous, $k_u \ge k$ holds for any $(\wt{{\bi r}}_u, k_u) \in \wt{T}$.

Indices $1 \le u  \le n$ of $T$ and indices $1 \le  u  \le n'$  of $\wt{T}$ are referred to as the {\em user identity} and {\em anonymized identity}, respectively.
The onto mapping from a user identity to an anonymized identity is defined by $\sigma: U \mapsto U'$, where $U$ and $U'$ are the domains of the user identities and the anonymized identities, respectively.

In recommendation systems, the number of items is generally large, which means that the rating vectors tend to have high dimensionality. 
Since the number of combinations of rating values exponentially increases with respect to the number of items, $k$-anonymization by means of generalization or suppression would seriously destroy the nature of the original rating values. 
For $k$-anonymization of high-dimensional numerical values, it is known that clustering or microaggregation preserves the utility of the original data. 
Here is an outline of $k$-anonymization by clustering: First, an algorithm clusters the vectors so that every cluster contains at least $k$ vectors. Then, each vector is replaced with the prototype of the cluster to which that vector is assigned. 
Algorithms for $k$-anonymization by clustering include the {one-pass $K$-means algorithm (OKA)\cite{Lin2008} and $r$-gather clustering \cite{Aggarwal2006}.
The recommendation models introduced in the following sections are not dependent on a specific anonymization method, but we used the OKA in our experiments.

\section{Recommendation Models Based on Anonymized Ratings}

In this section, we survey the available recommendation models that are based on anonymized ratings and then examine the risks of de-anonymization of the ratings.

\subsection{Stakeholders}\label{sec:stakeholder}

We first introduce the four stakeholders that occur in recommendations with anonymized ratings: the {\em rater}, the {\em rating collector}, the {\em recommender}, and the {\em user}.

The rater is the entity that gives the rating values to the rating collector.
The rating collector ({\em collector} for short) is the entity that collects the rating values from the raters and constructs the sparse rating matrix ${\bf R}$. Then, if necessary, the rating collector anonymizes the rating matrix as ${\wt{\bf R}}$ and distributes it to the recommender.
The recommender is the entity that obtains rating matrix ${\bf R}$ (or anonymized rating matrix ${\wt{\bf R}}$) from the collector and constructs the prediction function $f$. Then, upon request, the recommender provides recommendations to the users.

\subsection{Training Input}

The rating matrix that the recommender obtains from the collector is called the {\em training input}.
Two types of training inputs are considered in our models.

{\bf Case 1.} Let $U$ and $U'$ be the set of raters and users, respectively.
We assume $U' \subseteq U$, as in regular recommendation systems.
In this case, the prediction for any user can be personalized by using the ratings given by the user in the past. We represent the rating matrix in this case as ${\bi R}_+$ (Table \ref{tbl:model}, line 3).
If the collector anonymizes the rating matrix before providing it to the recommender, we call this Case 1A. In this case, the training input, i.e., the anonymized rating matrix, is represented as $\wt{{\bi R}}_+$ (Table \ref{tbl:model}, line 4).

{\bf Case 2.} 
In Case 2, we assume $U \cap U' = \emptyset$.
That is, none of the users who wish to obtain recommendations have previously given ratings to the collector. In this case, the predictions cannot be personalized based on past ratings. This situation is known as a {\em cold start}. The rating matrix in this case is represented by ${\bi R}_-$ (Table \ref{tbl:model}, line 5). If the collector provides the anonymized rating matrix $\wt{{\bi R}}_-$ of  ${\bi R}_-$, the situation is referred to as Case 2A (Table \ref{tbl:model}, line 6).

If the collector is a company whose customer base is comprehensive (e.g., a railway or cell phone provider), and the recommender wishes to provide recommendation services based on anonymized ratings purchased from the collector, then the recommendation system can be modeled with Case 1A.
However, if the ratings are collected from a limited segment of customers and are anonymized prior to distribution, then the recommendation system can be modeled with Case 2A; this is because the customer bases of the collector and the recommender are disjoint. 
Of course, there are intermediate situations between Case 1/1A and Case 2/2A that can be considered; however, we will not pursue these, because they can be covered by extensions of the existing models.

\subsection{Prediction Input and Models of Recommendation}

In order to provide a prediction that is personalized for a particular user, the recommender needs to have information about the user. This information is referred to as {\em prediction input}. Let $\omega(u)$ be the prediction input for user $u$. Below, we list possible variations of the prediction inputs. Table \ref{tbl:model} summarizes the relationships between the training input and prediction input for the different recommendation models.

\begin{enumerate}
\item User identity $\omega(u)=u$ as prediction input

Regular recommendations based on non-anonymized ratings are modeled with this prediction input; users provide their identities (Case1/REG). 
In Case 2 and Case 2A, user identities are not contained in the training inputs. 
In Case 1A, the recommender cannot know the relationship between the users and the anonymous raters of the training inputs.
Thus, in these cases, user identities cannot be used as prediction input.

\item Anonymous identity $\omega(u)=\sigma(u)$ as prediction input

Note that this prediction input can be used only in Case 1A, because in Case 2A, we assumed the ratings of users are not contained in the training input. 
In Case 1A, if an anonymous identity is given to the recommender as a prediction input, the prediction is personalized not for the user, but for the anonymous identity that contains the user. Because the prediction is personalized for the $k$ or more users that are contained in the anonymous identity, the effect of personalization can be weakened.

\item User ratings $\omega(u)={\bi r}_u$ as prediction input

In this case, users provide some of the ratings as prediction input.
In Cases 1A, 2, and 2A, the recommender cannot connect the prediction inputs with the raters in the training inputs. 
However, if a user can independently provide rating values, aside from the training inputs, the recommender can personalize the prediction for the user, based on the ratings provided (Case1A/UR, Case2/UR, and Case2A/UR). 
In Case1A/UR and Case2A/UR, the training inputs that the recommender obtains are anonymized, although the users provided non-anonymized ratings. If the users put more focus on the prediction accuracy of recommendation than on preserving anonymity, this decision is reasonable.

\item No prediction input: $\omega(u)=\emptyset$.

Without prediction input associated with the users, the predictions cannot be personalized. In this case, the recommender can do nothing but to give the average ratings of the items (BASELINE). 
\end{enumerate}

\begin{table*}[!t]
\caption{Recommendation models based on anonymized ratings}\label{tbl:model}
\begin{center}
\begin{tabular}{c|c|cccc}
\hline\hline
 & \raisebox{-1.8ex}[0pt][0pt]{training input} & \multicolumn{4}{|c}{prediction input} \\
\cline{3-6}
 &  & user id $u$ & user ratings $\bi r_u$ & anonymized id $\sigma(u)$ &  $\emptyset$ \\
\hline
Case 1 & $\bi R_+$ & Case1/REG & --- & --- &  \\
Case 1A & $\wt{\bi R}_+$ & --- & Case1A/UR & Case1A/AI & \raisebox{-1.8ex}[0pt][0pt]{BASELINE} \\
Case 2 & $\bi R_-$ & --- & Case2/UR & --- &  \\
Case 2A & $\wt{\bi R}_-$ & --- & Case2A/UR & --- &  \\
\hline
\end{tabular}
\end{center}
\end{table*}

\subsection{Risk of Privacy Leakage}\label{sec:risk}

When making recommendations with anonymized ratings, two types of privacy risks should be considered. 
In the first case, when the anonymized ratings are used as the training input, the recommender may try to use the anonymized rating vectors to identify the raters. 
In the second case, when the recommender obtains the prediction input from the users, the recommender may try to use the vectors to identify the users.

In the first case, the privacy risk is dependent on the guarantee of anonymity of the training input. If it is $k$-anonymous, the probability with which the recommender can identify the rater is at most $1/k$.

Next, we consider the risks in the second case.
For Case 2A, user ratings are not contained in the collection for any type of prediction input; there is thus no risk of reidentification.
For Case 1A/AI, the recommender obtains $\wt{\bi{R}}_+$ as the training input and the anonymized identity $\omega(u)=\sigma(u)$ as the prediction input. However, what the recommender can infer from this is no more than what the recommender could estimate from $\wt{\bi{R}}_+$ alone. Thus, the probability with which the recommender can reidentify the rater is again $1/k$ at most.

On the other hand, for Case1A/UR, if the recommender obtains $\omega(u)={\bi r}_u$ as the prediction input, the degree of anonymity can be decreased. For example, suppose the training input is a three-anonymized rating collection and Users 1, 2, and 3 all belong to the same anonymized identity. If the ratings of User 1 ${\bi r}_1$ are used as the prediction input, the recommender will be able to learn the anonymized identity to which User 1 belongs. The contribution of User 1 can then be removed from the rating vector $\sigma(1)$, which is given as the cluster center of the three users. This means that the guarantee of $k$-anonymity has been degraded from $k$-anonymity to $(k-1)$-anonymity.

Note that the anonymity can be degraded even if no information is provided that is associated with User 2 or 3.
For Case1A/UR, in order to guarantee $k$-anonymity after the prediction, the collector needs to anonymize the rating collection with some integer larger than $k$. In addition, the number of user ratings that the recommender can obtain needs to be controlled so that the $k$-anonymity of the training input is not compromised.

\section{Item-similarity Based Collaborative Filtering with Anonymized Ratings}

In this section, we present recommendation algorithms that use ratings anonymized by item-similarity based collaborative filtering\cite{Sarwar2001,Resnick1994}, using the models derived in the previous section.
All there algorithms use training inputs to construct groups of similar items prior to generating personalized predictions.

\subsubsection{Case1/REG}

We first introduce item-similarity based collaborative filtering without anonymization. 
The similarity of two items was defined to be the Pearson correlation coefficient of their ratings. Let $U_i$ be the set of users who rate item $i$. 
Then, the average rating of item $i$ and the correlation coefficient
between item $i$ and item $j$ are given by
\begin{align}
 r_{*i} &= \frac{1}{|U_i|} \sum_{u \in U_i} r_{u i}, \mbox{\hspace{0.5cm}}
 s_{ij} =
 \frac{\sum_{\ell = 1}^{m} (r_{\ell i} - r_{*i})(r_{\ell j} - r_{*j})}{
 \sqrt{\sum_{\ell = 1}^{m} (r_{\ell i} - r_{*j})^2}
 \sqrt{\sum_{\ell = 1}^{m} (r_{\ell j} - r_{*j})^2}}. \label{eq:sim}
\end{align}

With these item similarities, the predicted rating of item $i$ for user $u$ is given by
\begin{align}
 f(u,i;\omega(u)=u) &= r_{*i} + \frac{\sum_{\ell \in I_u} s_{i \ell} \left( r_{u\ell} - r_{*\ell} \right)}{
 \sum_{\ell \in I_u} |s_{i \ell}| },  \label{eq:simPP}
\end{align}
where $I_u$ is the set of items rated by user $u$.

\subsubsection{Case2/UR}

Let $u'$ be a user who wishes to obtain a recommendation. 
In the Case2/UR, the ratings of user $u'$ are not contained in ${\bi R}_-$. Since $I_{u'}=\emptyset$ in this case, it cannot be predicted by using eq.\ \ref{eq:simPP}. 
In order to personalize the predictions, users provide the ${\bi r}_{u'}$ of some of items as the prediction input. 
Noting that item similarities can be evaluated independent of users, the prediction for $u'$ can be given using ${\bi r}_{u'}$ as
\begin{align}
 f(u',i;\omega(u')={\bf r}_{u'}) &= r_{*i} + \frac{\sum_{\ell \in \bar{I}_{u'}} s_{i \ell} \left( r_{u'\ell} - r_{*\ell} \right)}{
 \sum_{\ell \in \bar{I}_{u'}} |s_{i \ell}| },  \label{eq:simPP2}
\end{align}
where $\bar{I}_{u'}$ is the set of items rated by user $u'$ in ${\bi r}_{u'}$. 
Here, note that $\bar{I}_{u'}$ can be arbitrarily chosen by user $u'$.

\subsubsection{Case1A/UR and Case2A/UR}

We now describe the algorithm for Case1A/UR.
The item similarities are evaluated from the training input $\wt{\bi R}_+ = (\wt{r}_{ui})$ by
\begin{align}
 \tilde{s}_{ij} &=
 \frac{\sum_{\ell = 1}^{m} (\tilde{r}_{\ell i} -
\tilde{r}_{*i})(\tilde{r}_{\ell j} - \tilde{r}_{*j})}{
 \sqrt{\sum_{\ell = 1}^{m} (\tilde{r}_{\ell i} -
\tilde{r}_{*i})^2}
 \sqrt{\sum_{\ell = 1}^{m} (\tilde{r}_{\ell j} -
\tilde{r}_{*j})^2}}, \label{eq:sim-anonymize}
\end{align}
where $\tilde{r}_{*i}$ is the average rating of item $i$ evaluated with $\wt{\bi R}_+$.

Then, the prediction is made using the non-anonymized user ratings given as prediction inputs by
\begin{align}
 f(u,i;\omega(u)={\bi r}_u) &= \wt{r}_{*i} + \frac{\sum_{\ell \in \bar{I}_u} \wt{s}_{i \ell} \left( r_{u\ell} -
\wt{r}_{*\ell} \right)}{
 \sum_{\ell \in \bar{I}_u} |\wt{s}_{u \ell}| }.  \label{eq:simAoP}
\end{align}
Prediction by eq.\ \ref{simAP} makes use of the average item ratings and the item similarities estimated from the anonymized rating matrix, whereas the prediction is made for a particular user. By giving user ratings as prediction inputs, the prediction can be personalized even when the rating matrix has been anonymized.

In the algorithm for Case2A/UR, the similarities can be obtained by using eq.\ \ref{eq:sim-anonymize}.
The predictions can be made with eq.\ \ref{eq:simAoP} if $\bar{I}_u$ is replaced with $\bar{I}_{u'}$, as in eq.\ \ref{eq:simPP2}.

\subsubsection{Case1A/AI}

In this case, the similarities can be evaluated by eq.\ \ref{eq:simAoP}.
The prediction of the rating of item $i$ for user $u$ becomes
\begin{align}
 f(u,i;\omega(u)=\sigma(u)) &= \wt{r}_{*i} + \frac{\sum_{\ell \in \wt{I}_{\sigma(u)}} \wt{s}_{i \ell} \left(
\wt{r}_{\sigma(u)\ell} - \wt{r}_{*\ell} \right)}{
 \sum_{\ell \in \wt{I}_{\sigma(u)}} |\wt{s}_{i \ell}| },  \label{eq:simAP}
\end{align}
where $\wt{I}_{\sigma(u)}$ is the set of items rated by anonymous identity $\sigma(u)$ in the anonymized rating matrix $\wt{\bi{R}}_+$.
Note that predictions made by this equation are personalized for an anonymous user identity $\sigma(u)$, not for a particular user $u$.

\subsubsection{BASELINE}

In the BASELINE model, the recommender does not have any information that can be used to personalize the prediction. Therefore, the prediction is the average of the anonymized ratings of the specified item: $r_{*i}$ as evaluated by eq.\ \ref{eq:sim}.

In summary, Case1/REG is equivalent to \cite{Sarwar2001}. Case2/UR is a well-known extension of Case1/REG for the cold-start setting. Case1A/AI is a model similar to \cite{polat2003privacy} or \cite{chang2010towards}.
Case1A/UR and Case2A/UR are models that, to the best of our knowledge, are newly introduced in this paper.

\section{Experiments}\label{sec:experiment}

In this section, we report the results of experiments carried out in order to evaluate the effects of anonymization on the prediction accuracy of recommendations. 
For these experiments, we used the MovieLens datasets\cite{Resnick1994}.
The 100k dataset contains 100,000 ratings of 1682 items given by 943 users.
The 1M dataset contains about a million ratings of 3883 items given by 6040 users.

\subsection{Experiments for Case 1 and Case1A}

\begin{figure*}[!t]
 \centering
\includegraphics[scale = 0.25]{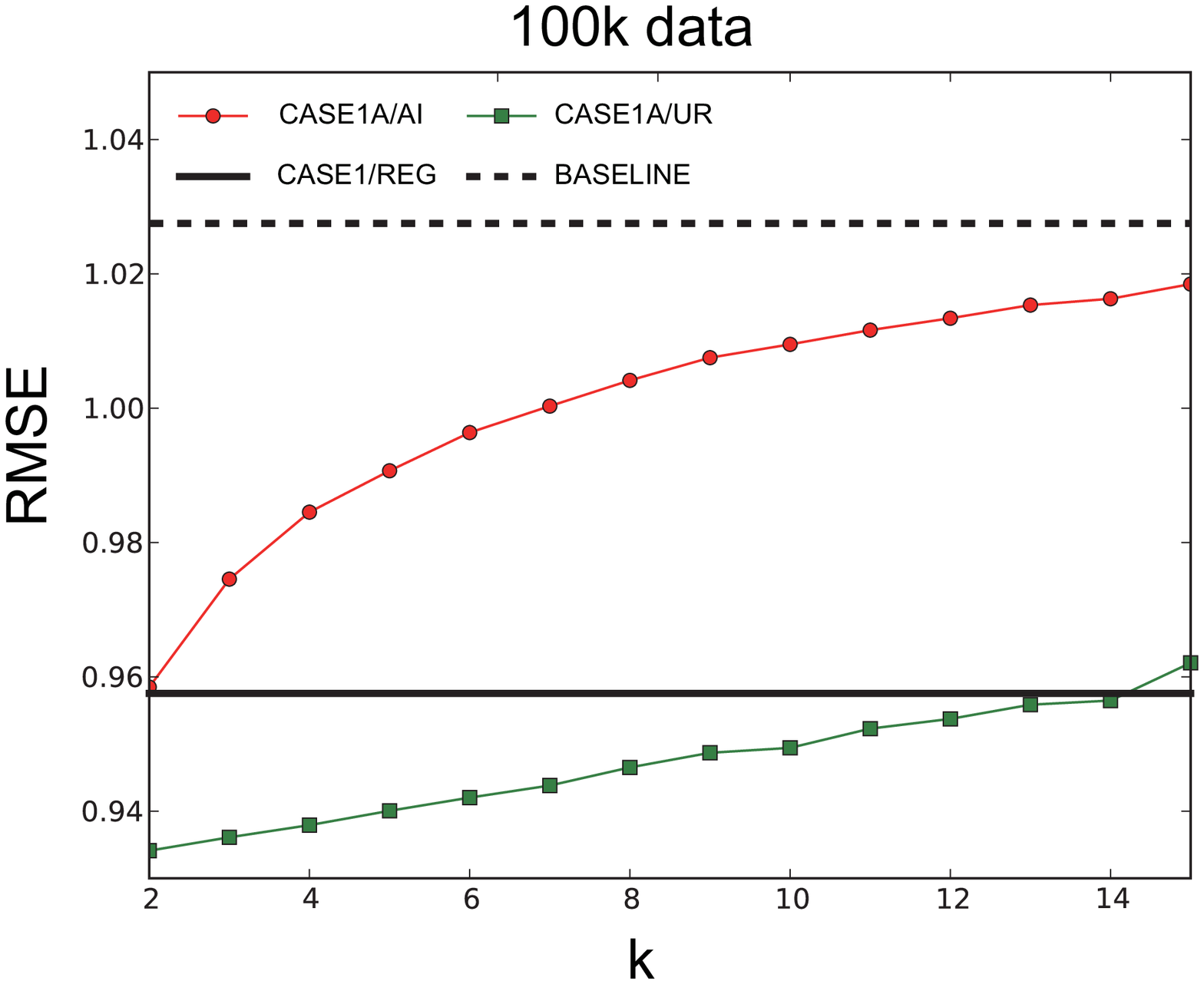}
\includegraphics[scale = 0.25]{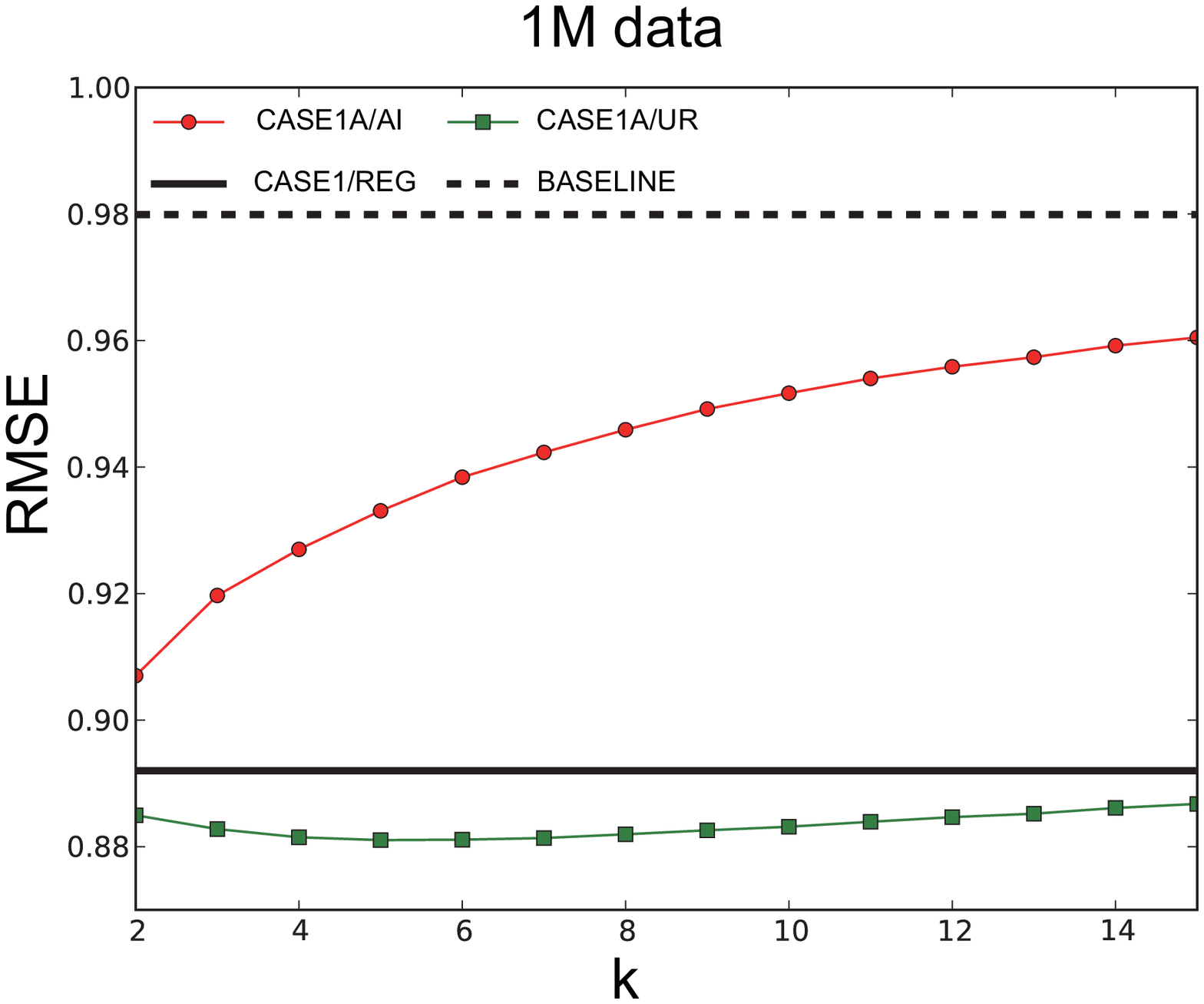}
 \caption{Change of RMSE w.r.t.\ $k$ for Case1/REG, Case1/UR, Case1/AI, and the BASELINE model. Left: 100k data, Right: 1M data.
 \label{fig:result-similarity}}
\end{figure*}

\subsubsection{Settings}

For Case 1 and Case 1A, we assumed that all the users who received personalized predictions were contained in the set of raters, i.e., $U' \subseteq U$. 
For Case1/REG, we uniformly chose 80\% of the ratings from $\bi R$ at random and used them as the training input. The remaining ratings were used for the test data to evaluate the RMSE.

For Case1A/AI and Case1A/UR, 80\% of the ratings from $\bi R_+$ were chosen uniformly at random, and then these ratings were $k$-anonymized using the OKA method\cite{Lin2008}; these were used as the training input. The remaining 20\% of the ratings were used as the test data. 
For Case1A/UR, 20\% of the non-anonymized ratings ${\bi r}_u$, chosen uniformly at random from the training data, were used for the prediction. 
In the experiments, the RMSE of the recommendations with these models were measured while varying the anonymity parameter as $k=2, 3,\hdots, 15$. 

\subsubsection{Results}

The change of in the RMSE w.r.t.\ the anonymity parameter $k$ for Case1/REG, Case1A/UR, Case1A/AI, and the BASELINE are shown in Fig.\ \ref{fig:result-similarity}} . 
The RMSEs of Case1/REG are less than those of Case1/AI for all evaluated values of $k$. 
For Case1/AI with a larger $k$, the anonymous identities contain a larger number of individuals; this makes the target of the personalization vague, and the prediction accuracy deteriorates. 
This indicates that there is a trade-off between utility and privacy for Case1/AI.

In Case1/UR, the RMSE deteriorates as $k$ increases. 
However, surprisingly, Case1A/UR achieves a prediction accuracy that is better than that of Case1/REG for all evaluated values of $k$. 
This behavior is further discussed in Section  \ref{sec:fin}. 
We must bear in mind that the anonymity in Case1A/UR can be weakened if the user ratings are given to the recommender as prediction input, as discussed in Section \ref{sec:risk}. 
Because the levels of anonymity achieved for Case1A/AI and Case1A/UR are not equivalent, the RMSE of Case1A/AI and Case1A/UR are not directly comparable. 

For Case 1A/UR, the results are significantly different between the 1M dataset and the 100k dataset.
With the 1M data, the RMSE slightly improves (decreases) as $k$ increases in the range of $2 \le k \le 5$, then the RMSE increases as $k$ increases in the range of $5 < k \le 15$. 
This behavior is further discussed (and further experimental results are presented) in Section \ref{sec:fin}.

\subsection{Experiments for Case 2 and Case 2A}

\begin{figure}[!t]
 \centering
\includegraphics[scale = 0.27]{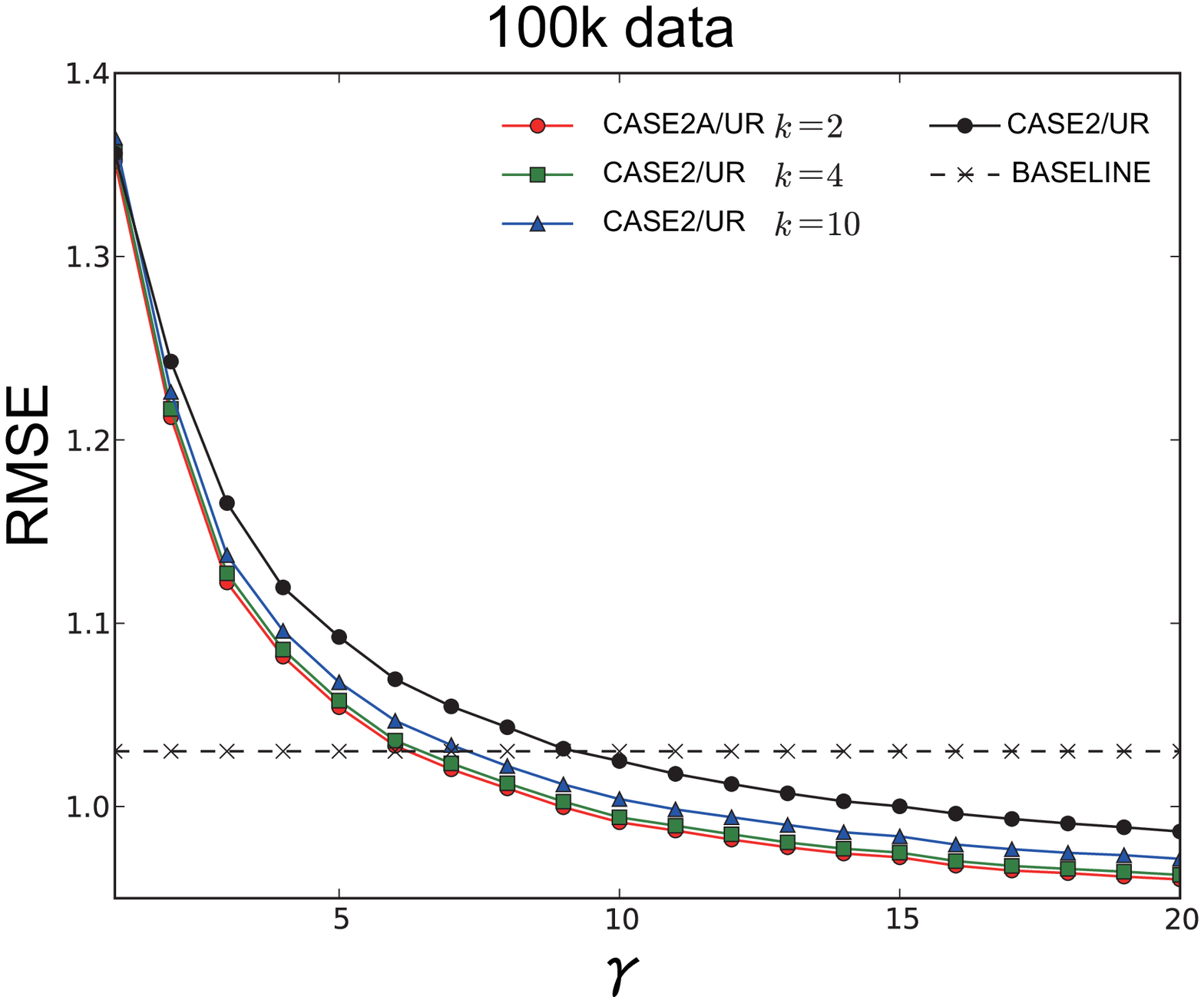}
\includegraphics[scale = 0.27]{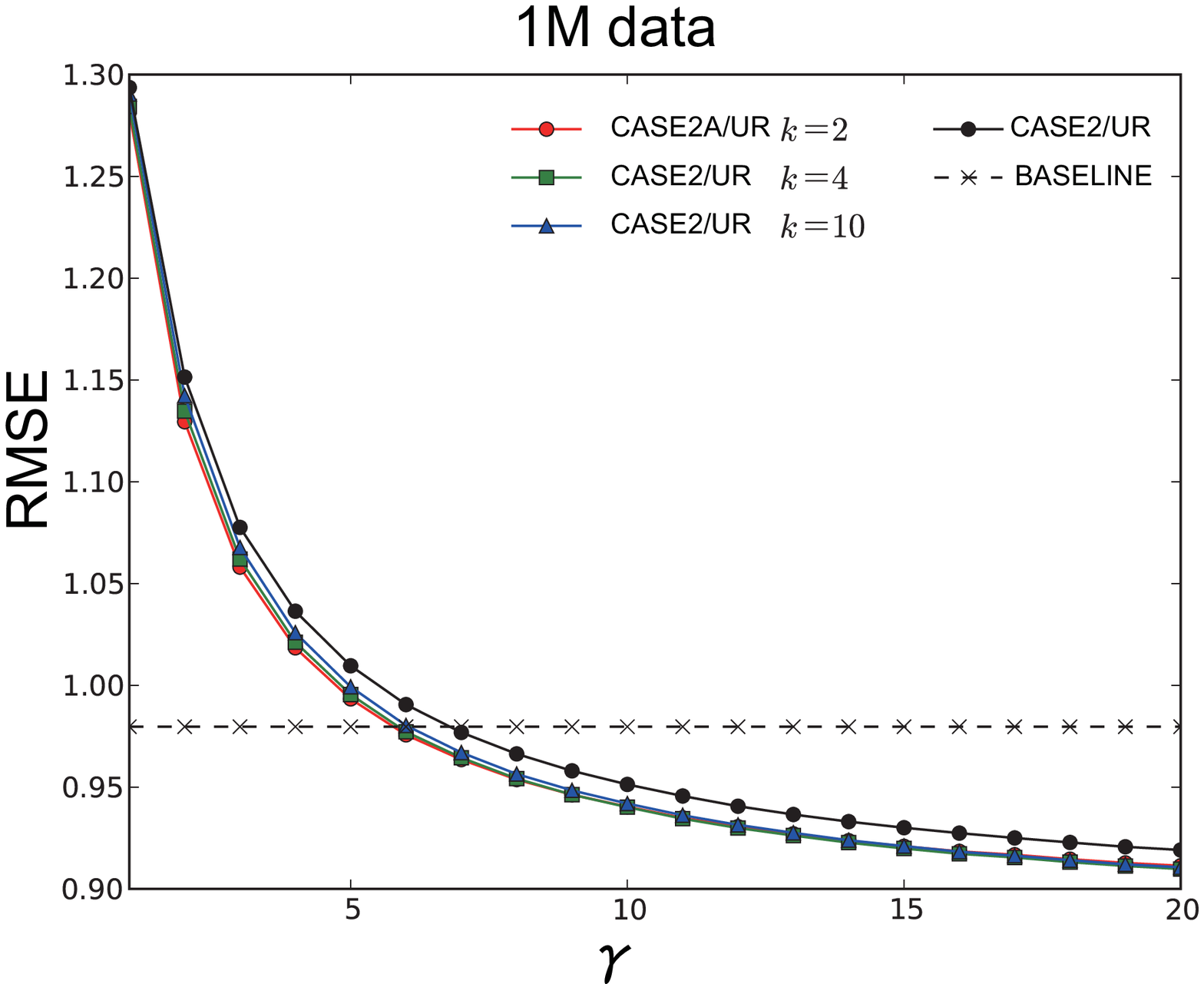}
 \caption{Change of RMSE w.r.t.\ \num{}, the number of user ratings used as the prediction input, for Case 2 and Case 2A. Case2/REG and Case2A/UR for $k=2,4$, and $10$, and the BASELINE are compared. Left: 100k data, Right: 1M data.
 \label{fig:result-case3}}
\end{figure}

\subsubsection{Settings}

For Case 2 and Case 2A, we assumed a cold start; none of the users who received personalized predictions were contained in the set of raters, i.e., $U' \cap U= \emptyset$. 
In Case2/UR, we uniformly chose 80\% of the users in $U$ at random, and their ratings were used as the training inputs. The ratings of the rest of the users were used for the test data to evaluate the RMSE.
For the remaining 20\% of the users,  \num{} ratings were used as their prediction inputs and the rest of ratings were used as the test data to evaluate the RMSE. To evaluate the RMSE, $20$ \num{} ratings were chosen uniformly at random to be used as prediction inputs; the RMSE averaged over these $20$ trials was used as the RMSE.
In these experiments, the number of ratings used as the prediction inputs, \num{}, was varied from $1$ to $20$; the anonymity parameter was set to $k=2,4,10$. With these settings, we compared the changes in the RMSE for the BASELINE and the Case2/UR and Case2A/UR models w.r.t.\ \num{}.

\subsubsection{Results}

Figure \ref{fig:result-case3} shows the changes in the RMSE of the BASELINE, Case2/REG, and Case2A/UR w.r.t.\ \num{}.
Case2A/UR had a better the prediction accuracy than did Case2/UR, for all values of $k$ and \num{}.
From these results, we can see that item-similarity matrices evaluated with anonymized ratings gave better predictions for both Case 1A/UR and Case 2A/UR. 
Furthermore, we can see that the RMSEs are not greatly affected by the degree of anonymity $k$ for Case2A/UR. 
Thus, if $k \le 10$, anonymization of the ratings does not damage the recommendations.

Another important observation is that the RMSE improves as \num{}, the number of ratings used in the prediction input, increases.
This is because the prediction can be better personalized when there are more user ratings used as prediction inputs.
The RMSE of Case2A/UR is better than that of the BASELINE when \num{}$ \ge 6$ in the 1M dataset and \num{}$> \ge 8$ in the 100k dataset.
Thus, a practical solution for giving recommendations with anonymized ratings is for the users to rate only a very small number out of thousands of items. 
Note that the ratings used for the prediction inputs can be arbitrarily chosen by the users. 
If the users provide ratings of popular items that are rated by a large number of users, any privacy leakage caused by giving prediction inputs to the recommender can be ignored.

\subsection{Analysis of Similarity Matrix Evaluated with Anonymized Ratings}\label{sec:fin}

From the experimental results in the previous subsections, we observe that the prediction accuracy of recommendations based on anonymized ratings can be better than those based on non-anonymized ratings in some settings; thus contradicts our intuition.
In this subsection, we discuss the reasons of these useful but counterintuitive results.

The ratings matrices used for the recommendations are usually quite sparse. In the Movielens dataset, the sparsity is 6.3\% in the 100k dataset and 4.2\% in the 1M dataset. 
In our experiments, clustering (OKA) was used for anonymization, and the anonymized user ratings were set to the average of the user ratings that belonged to the identical anonymous identities.
If some but not all users who belonged to an anonymous identity rated an item, when the ratings were anonymized, the ratings of all members of the identity were set to the average rating of those users who did rate the item. 
Because of this manipulation, as $k$ increases, a larger number of unrated elements were complemented with the average ratings. Thus, the anonymized rating matrix became denser. 
On the one hand, if the rating matrix became denser because $k$ was larger, the item similarities can be estimated with a larger number of ratings, and this might make the estimation of the similarities more precise. 
On the other hand, when $k$ is larger, the number of users contained in a single anonymous identity also becomes larger. Thus, from the perspective of the users, the ratings of the anonymized identity do not accurately reflect their individual ratings, and this might cause the prediction accuracy to deteriorate.
Thus, there exists a dilemma in setting the value of $k$. 

When recommendations are based on item similarities, predictions are made based on the average ratings of the items and the average of the user ratings weighted by the item similarities. 
In order to see how $k$ affects the prediction accuracy, we examined the behaviors of the average ratings of the items and the item similarities w.r.t.\ changes in $k$.

\begin{figure*}[!t]
 \centering
\includegraphics[scale = 0.19]{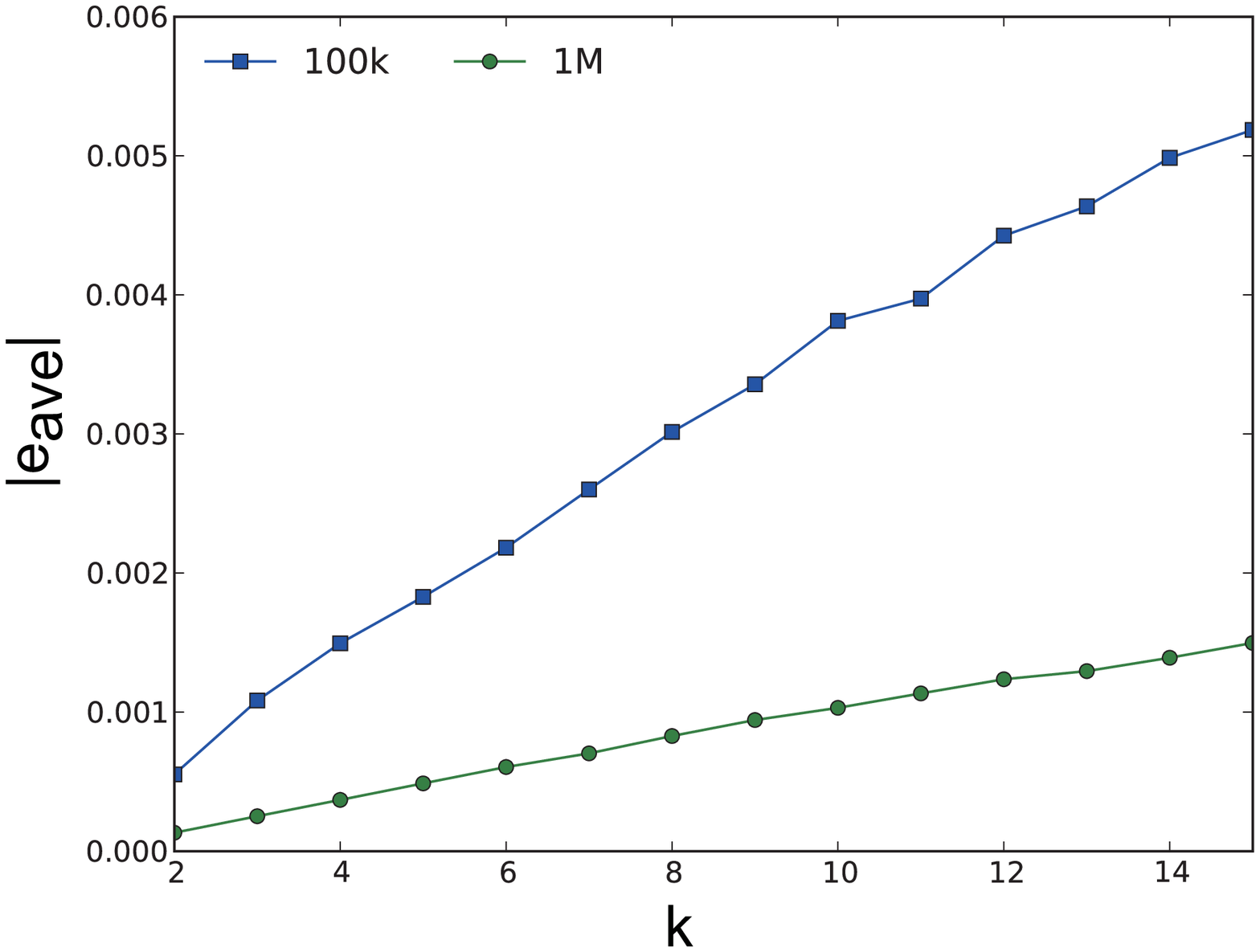}
\includegraphics[scale = 0.19]{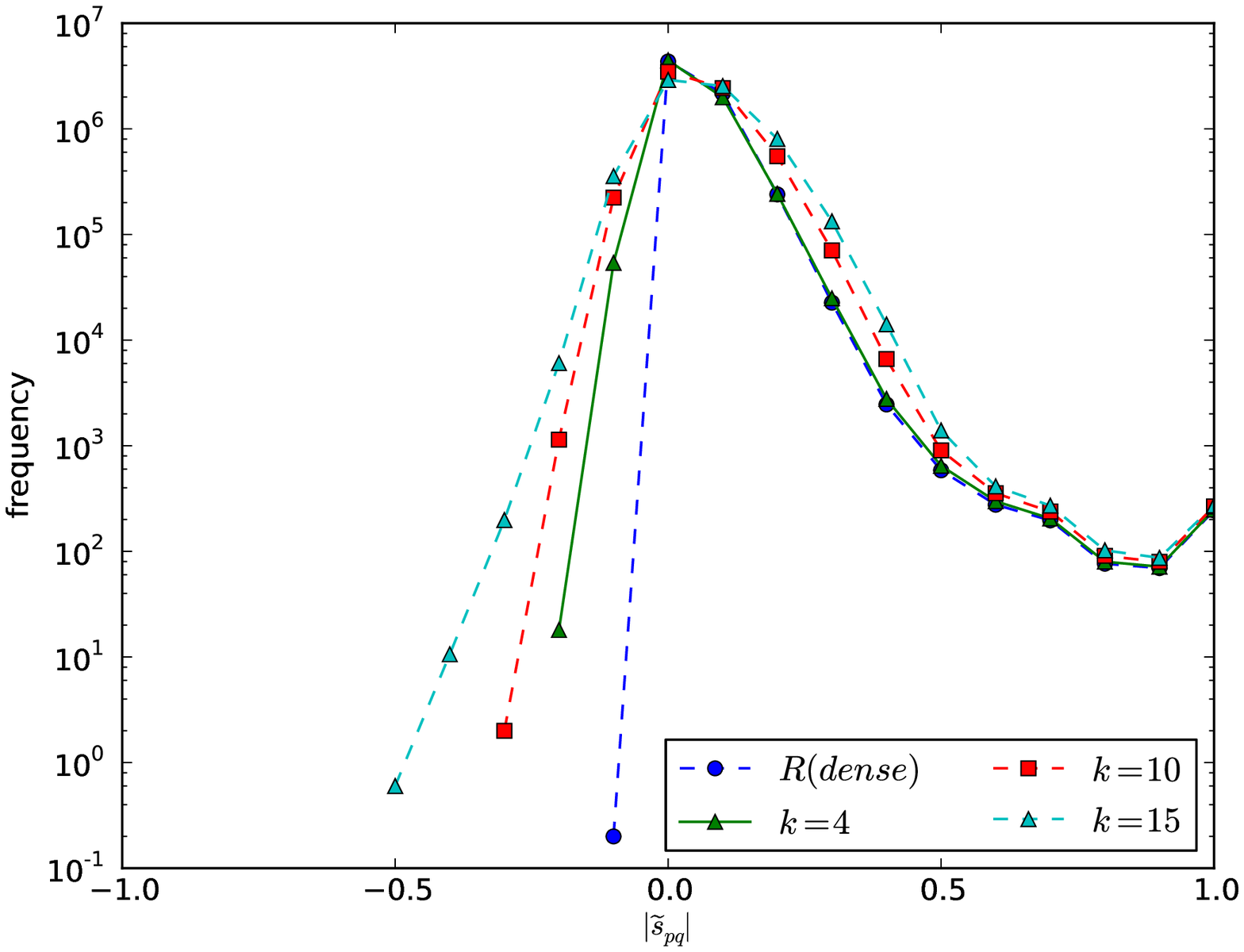}
\includegraphics[scale = 0.19]{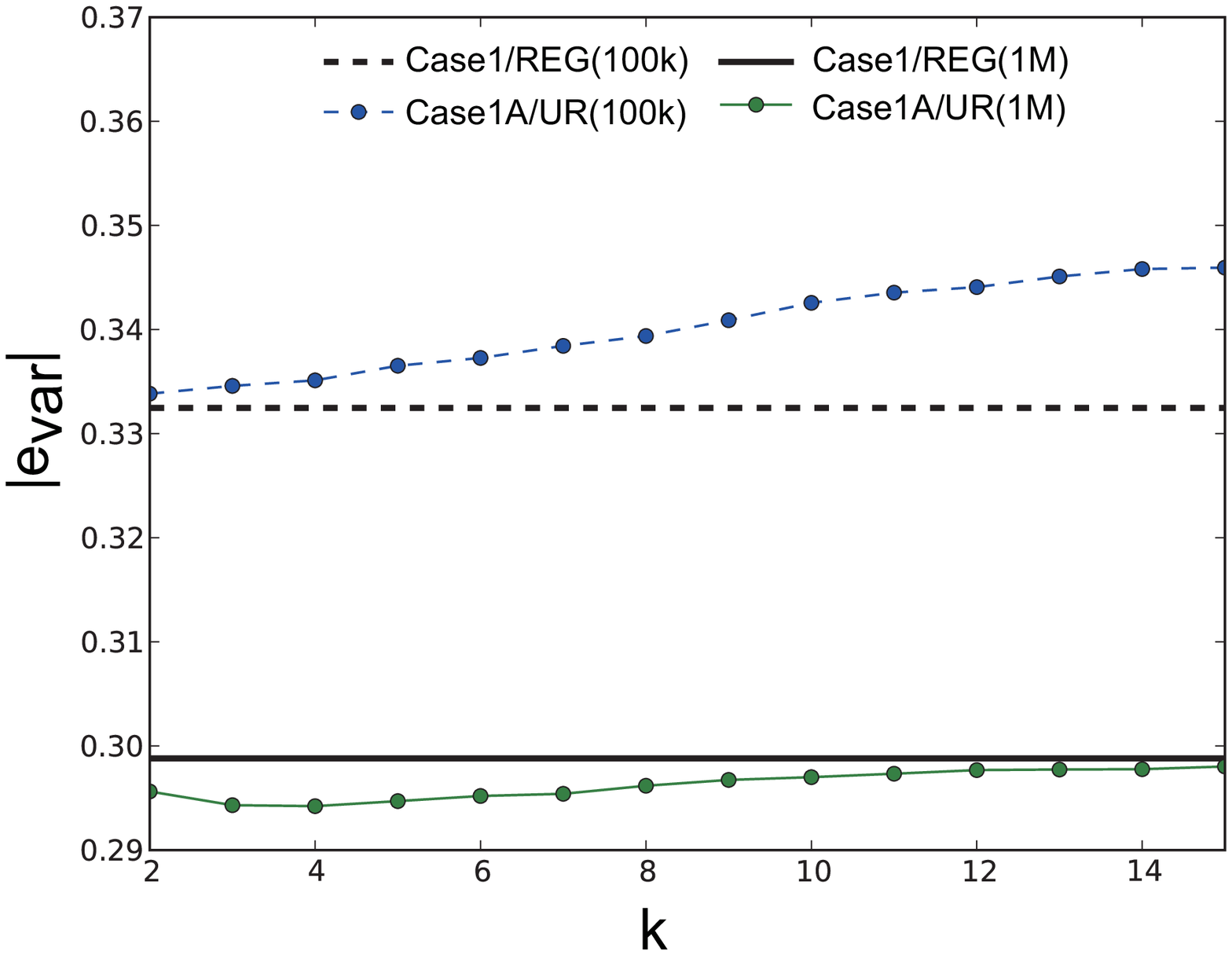}
 \caption{(a) Change of $e_{\mbox{ave}}$, the difference between the average ratings of items in $\bi{R}$ and $\wt{\bi{R}}$, w.r.t.\ $k$; (b) Histogram of $\wt{s}_{ij}$; (c) Change of $e_{\mbox{var}}$, the variance of the difference between rating $r_{ij}$ and the predicted rating $\hat{r}_{ij}$, w.r.t.\ $k$.}
 \label{fig:item-average}
\end{figure*}

Figure \ref{fig:item-average}(a) shows the changes in the item ratings $e_{\mbox{avg}}=\frac{1}{|M|}\sum_{i=1}^{M}|  r_{*i} - \tilde{r}_{*i}|$.
In both the 100k dataset and the 1M dataset, $e_{\mbox{avg}}$ becomes larger as $k$ becomes larger; however, the absolute value of the error is kept within $0.01$: the anonymity degree $k$ does not greatly affect the average ratings of the items.

Figure \ref{fig:item-average}(b) shows the histogram of the item similarities $\wt{s}_{ij}$ with different $k$ values.
From the figure, we can see that the positive similarities frequently appear in the non-anonymized item-similarity matrix, while the negative similarities do not. 
This tendency changes with larger $k$. More precisely, by complementing the sparse rating matrix by anonymization with larger $k$, the frequency of negative similarities increases, whereas that of the positive similarities does not change. 
The estimation of item similarities between two items by using eq.\ \ref{} becomes more accurate as the number of users who rate both of the items increases. 
Thus, the results shown in Fig.\ \ref{fig:item-average}(b) indicate that anonymization enhances the evaluation of the item similarities between items that are not similar to each other.
We should bear in mind that the similarity estimation with anonymized rating matrices does not necessarily provide ``better'' similarities compared to those estimated with non-anonymized rating matrices.

Finally, we consider the change in the variance of the prediction: $e_{\mbox{var}} = |r_{ui} - \hat{r}_{ui}|$ w.r.t.\ $k$, where $\hat{r}_{ui}$ is the prediction of the rating of item $i$ for user $u$.
In Fig.\ \ref{fig:item-average}(c), the changes of variance $e_{\mbox{var}}$ are shown for the BASELINE, Case1/REG, Case1A/UR, and Case1A/AI.

For Case1/AI, the variance increases with $k$. 
The prediction for Case1/AI is personalized w.r.t.\ the anonymous identities. 
In this situation, the prediction might not be appropriately personalized when $k$ is large, because the number of users per anonymized identity also becomes large.

For Case1/UR, the variance decreases from $k=2$ to $5$, then it increases. Note that the variances for Case1/UR are never larger than those for Case1/REG. 
For Case1/UR and Case1/REG, the user ratings used for prediction are the same, and only the similarities are different. Thus, these results indicate that the item similarities obtained from anonymized ratings are more robust than those obtained from non-anonymized ratings. 
For Case1/UR, when $k=4$, these are in balance and the best prediction accuracy is achieved.

\section{Conclusion}

In this paper, we considered recommender systems that used anonymized ratings and with various types of input for training the recommendation function and for personalization of the prediction. Then, based on these models, we presented item-based collaborative filtering algorithms for providing recommendations. 
Our experimental results show that item-based collaborative filtering performs better with anonymized ratings than with non-anonymized ratings, when the users show 5--10 (non-anonymized) ratings to the recommender. 
This surprising result indicates that privacy protection does not necessarily degrade the usefulness of recommendations.
From analysis of this counterintuitive result with experiments, we observed that the sparsity of the ratings can be reduced by anonymization, and the variance of the prediction can be reduced if $k$, the anonymization parameter, is appropriately tuned. Because of these effects, the predictive performance of the recommendation with anonymized ratings can be improved. 
Our future work is to expand our models to other recommendation algorithms, including matrix factorization.

\bibliographystyle{splncs03}
\bibliography{reference}

\end{document}